\documentclass[12pt]{article}
\usepackage{amsmath}
\usepackage{amssymb}
\usepackage{mathtext}
\usepackage{graphicx}

\begin{document}

\section*{STABILITY OF THOMSON'S \\
CONFIGURATIONS OF VORTICES ON A
SPHERE}\footnote{REGULAR AND
CHAOTIC DYNAMICS V.5, No. 2, 2000\\
{\it Received January 17, 2000\\
AMS MSC 76C05, 58F10}\bigskip}

\begin{centering}
A.\,V.\,BORISOV\\
Faculty of Mechanics and Mathematics,\\
Department of Theoretical Mechanics
Moscow State University\\
Vorob'ievy gory, 119899 Moscow, Russia\\
E-mail: borisov@uni.udm.ru\medskip\\
A.\,A.\,KILIN\\
Laboratory of Dynamical Chaos and Non Linearity,\\
Udmurt State University\\
Universitetskaya, 1, Izhevsk, Russia, 426034\\
E-mail: aka@uni.udm.ru\\
\end{centering}

\begin{abstract}
In this work stability of polygonal configurations
on a plane and sphere is investigated.
The conditions of linear stability are obtained.
A nonlinear analysis of the problem is made with the help of Birkhoff
normalization.
Some problems are also formulated.
\end{abstract}

\section{A linear stability of Thomson's configurations
of vortices on a sphere}

Motion equations of a system of $N$ point vortices on a
sphere can be presented in the
Hamiltonian form~\cite{Bgm} with the Poisson bracket
\begin{equation}
\label{f1}
\{\varphi_i,\,\cos\theta_k\}=\frac{\delta_{ik}}{R^2\Gamma_i}
\end{equation}
and the Hamiltonian
\begin{equation}
\begin{split}
\label{f2}
H=-\frac{1}{8\pi}\mathop{{\sum}'}\limits_{i,\,k=1}^{N}\Gamma_i
\Gamma_k\ln\left(4R^2\sin^2\frac{\gamma_{ik}}{2}\right)\mbox{, } \\
\cos\gamma_{ik}=\cos\theta_i\cos\theta_k+\sin\theta_i\sin\theta_k
\cos(\varphi_k-\varphi_i)\,,
\end{split}
\end{equation}
where~$\theta_i$, $\varphi_i$ and~$\Gamma_i$~are spherical
coordinates and the intensity of the~$i$-th vortex, $\gamma_{ik}$~is an angle
between the~$i$-th and the~$k$-th vortices, $R$~is a radius of the sphere
the prime denotes that $i\ne k$.
Here and further all indices values from~1 to~$N$.
Motion equations of such a system have the form
\begin{equation}
\label{f3}
\left\{
\begin{aligned}
\dot\theta_k & =-\frac{1}{4\pi R^2}\mathop{{\sum}'}\limits_{i=1}^{N}\Gamma_i
\frac{\sin\theta_i\sin(\varphi_k-\varphi_i)}{1-\cos\gamma_{ik}}\,, \\
\sin\theta_k\dot\varphi_k & = \frac{1}{4\pi R^2}\mathop{{\sum}'}\limits_{i=1}^{N}
\Gamma_i\frac{\sin\theta_k\cos\theta_i-\cos\theta_k\sin\theta_i
\cos(\varphi_k-\varphi_i)}{1-\cos\gamma_{ik}}\,.
\end{aligned}
\right.
\end{equation}

In the case of equal intensities~$\Gamma_i$ the system is invariant
with respect to the discrete group of permutations of vortices, and therefore
 allows different symmetric partial solutions. Let us consider
partial solutions, which are analogous to the
 well known Thomson's configurations on a plane. Then the vortices are
located on the same latitude~$\theta_0$ in the vertices of a regular~$N$-gon
and  rotate around its center with the angular velocity~$\omega$:
\begin{equation}
\label{f4}
\theta_i=\theta_0=const\,,\qquad \varphi_i=\omega(\theta_0)t+\frac{2\pi}{N}(i-1)\,,\qquad
\omega(\theta_0)=\frac{\Gamma(N-1)}{4\pi R^2}\frac{\cot\theta_0}{\sin\theta_0}\,.
\end{equation}

Let us consider stability of these partial solutions in linear
approximation. The study of stability of Thomson's configurations of
vortices on a plane in linear approximation was carried out by
J.\,Thomson~\cite{Tom}. He has shown that they are stable in linear
approximation for~$N\le7$, and unstable for~$N>7$.  Let us take
variations~$\delta\theta_i$ and~$\delta\varphi_i$ as
\begin{equation}
\label{f5}
\theta_i=\theta_0+\delta\theta_i\,,\qquad
\varphi_i=\omega(\theta_0)t+\frac{2\pi}{N}(i-1)+\delta\varphi_i\,.
\end{equation}
The linearized equations for~$\delta\theta_i$ and~$\delta\varphi_i$ are
autonomous and have the form:
\begin{equation}
\label{f6}
\left\{
\begin{aligned}
\delta\dot\theta_k & =\frac{\Gamma}{4\pi R^2}\frac{1}{2\sin\theta_0}A_{ki}
\delta\varphi_i\,, \\
\sin\theta_0\delta\dot\varphi_k & = \frac{\Gamma}{4\pi R^2}\frac{1}{2\sin^2
\theta_0}A_{ki}\delta\vartheta_i-\frac{\Gamma(N-1)}{4\pi R^2}\frac{1+\cos^2
\theta_0}{\sin^2\theta_0}\delta\theta_k\,,
\end{aligned}
\right.
\end{equation}
where~$A$~is an $(N\times N)$-matrix with the elements
\begin{equation}
\label{f7}
A_{ki}=\delta_{ki}\sum\limits_{m=1}^{N-1}\frac{1}{\sin^2\frac{\pi}{N}m}
-(1-\delta_{ki})\frac{1}{\sin^2\frac{\pi}{N}(k-i)}\,.
\end{equation}
By differentiating with respect to time, the equations~(\ref {f6}) can be reduced
to a system of $N$ second-order equations in the following form:
\begin{equation}
\label{f8}
\delta\ddot\varphi_k=\Biggl(\frac{\Gamma}{8\pi R^2\sin^2\theta_0}\Biggr)^2
\left(A_{kj}-2(N-1)(1+\cos^2\theta_0)\delta_{kj}\right)A_{ji}
\delta\varphi_i
\end{equation}
(similar equations are obtained for~$\theta_k$ under elimination
of~$\varphi_k$).

A solution of the equation~(\ref{f8}) has the form
$$
\delta\varphi_k=\sum\limits_m C^{(m)}_ke^{\Omega_mt},
$$
where the constants~$C_k^{(m)},\,\Omega_m$ are expressed through the
eigenvalues and eigenvectors of the matrix~$A_{ki}$.
As in the case of a plane~\cite{Aref}, the elements of~$A_{ki}$ depend
only on differences~$(k-i)$, thus the matrix~$A$ can be diagonalized
by the Fourier transformation, and the eigenvalues of the matrix can be
presented as
\begin{equation}
\label{f9}
\lambda_m=2m(N-m)\,, \quad m=0,\,1,\,\ldots,\,\Bigl[\frac{N}{2}\Bigr]\,.
\end{equation}
Each eigenvalue corresponds to two
eigenvectors~$\phi_i^{(m)}=\cos\left(\frac{2\pi m}{N}i\right)$
and~$\psi_i^{(m)}=\sin\left(\frac{2\pi m}{N}i\right)$, ${i=1,\,\ldots,\,N}$.
To the values~$m=0$ and~$m=\smash{\frac{N}{2}}$ (in case of even~$N$),
there corresponds a unique eigenvector~$\phi_i^{(m)}$ \cite{Aref}.

Using~(\ref{f8}),~(\ref {f9}), we find frequencies~$\Omega_m$
\begin{equation}
\label{f10}
\begin{aligned}
\Omega_m & =\frac{\pm\Gamma\lambda_m^{\frac{1}{2}}}
{8\pi R^2\sin^2\theta_0}\sqrt{\lambda_m-2(N-1)
(1+\cos^2\theta_0)}= \\
{} & =\frac{\pm\Gamma\sqrt{m(N-m)}}{4\pi R^2\sin^2\theta_0}
\sqrt{m(N-m)-(N-1)
(1+\cos^2\theta_0)}\,.
\end{aligned}
\end{equation}

The presence of the zero frequency~(\ref{f10}) at~$m=0$ corresponds to
instability of Thomson's configurations in the absolute coordinate system.
Indeed, under perturbations~$\delta\theta_i$, $\delta\phi_i$ corresponding
to~$m=0$ the Thomson's configuration~(\ref{f4}) transforms into
Thomson's configuration, which is close to
the latter and moves away from it linearly
with time.
Stability of relative motions of the vortices, from which we exclude the
rotations of the system as solid-state configuration, are determined by the
remained frequencies.
It is necessary for the relative stability
 the remained frequencies~$\Omega_m$ purely
imaginary. It is obvious from~(\ref{f10}) that
the configurations~(\ref{f4}) gradually lose their stability approaching the
equator (i.\,e. with magnification of~$\theta_0$).

Table 1 shows the boundaries of varying of the radicand in~(\ref{f10})
for various~$N$ with the maximum value~$m(N-m)$ and also values
of~$\theta_0^*$, at which the radicand changes the sign.

For~$N\ge7$ there is always at least one frequency with a positive real part,
hence, such configurations are unstable already in the linear approximation.

\begin{table}
\begin{center}
\begin{tabular}{l}
\tablename{\,1.} \\
\begin{tabular}{|c|c|c|c|c|}
\hline
{\footnotesize $N$} & {\footnotesize $\max(m(N-m))$} &
\multicolumn{2}{|c|}{\footnotesize $ m(N-m)-(N-1)(1+\cos^2\theta_0)$}
& {\footnotesize $\theta^*_0$} \\
\cline {3-4}
{} & & $\footnotesize\theta_0=0$ & $\footnotesize\theta_0=\frac{\pi}{2}$ & \\
\hline
2 & 1 & $-1$ & 0 & $\frac{\pi}{2}$ \\
\hline
3 & 2 & $-2$ & 0 & $\frac{\pi}{2}$ \\
\hline
4 & 4 & $-2$ & 1 & $\arccos\frac{1}{\sqrt{3}}$ \\
\hline
5 & 6 & $-2$ & 2 & $\frac{\pi}{4}$ \\
\hline
6 & 9 & $-1$ & 4 & $\arccos\frac{2}{\sqrt{5}}$ \\
\hline
7 & 12 & 0 & 6 & 0 \\
\hline
8 & 16 & 2 & 9 & --- \\
\hline
\end{tabular}
\end{tabular}
\end{center}
\vspace{-8mm}
\end{table}

For~$N=6,\,5,\,4,\,3$ and 2 and for~$\theta_0\in(0,\,\theta_0^*)$
all frequencies~$\Omega_m$ are purely imaginary.
Thus, under the above-mentioned
conditions the specified configurations are stable in
linear approximation. For~$ \theta_0\in [\theta_0^*,\,\frac{\pi}{2}]$ these
configurations are either unstable in linear approximation or
consideration of higher order
approximations (at~$\theta_0=\theta_0^*$) is required for determination of
stability.

\section[Nonlinear stability]{A nonlinear stability of Thomson's
configurations on a sphere and a plane for~$\boldsymbol{N =2,\,\ldots,\,6}$}

The work by L.\,G.\,Khazin~\cite{Hasin} introduced a proof of the Lyapunov
stability of Thomson's configurations in nonlinear statement on a plane.
However, though the general theorem from the stability theory, proved by
L.\,G.\,Khazin in this work is correct, its application to Thomson's
configurations on a plane is not appropriate. Zero frequencies considered
by Khazin correspond to the motion of vortices as a whole
and are not connected
with the relative stability of Thomson's configurations. Thus, the problem of
stability of Thomson's configurations on a plane in strict nonlinear
statement was not solved.

Let us show the Lyapunov stability of Thomson's configurations on a sphere
and plane relatively to the perturbations of mutual distances between
vortices (further we shall speak about ``relative'' stability)
for~$N=2,\,\ldots,\,6$.

\subsection*{The case of $\boldsymbol{N=2}$}

The most general kind of motion of two vortices on a sphere (plane) is the
uniform rotation around a motionless axis, which passes through the center
of vorticity and the center of sphere (perpendicular plane) with
preservation of the mutual distance~\cite{BorLeb,BorMam}. Under a small
perturbation in this system the distance between vortices changes a little,
but it is also preserved during the motion. Thus, Thomson's
configurations of two vortices are ``relatively"  Lyapunov stable both on
a sphere and on a plane.

\subsection*{The case of $\boldsymbol{N=3}$}

Let us consider a geometric interpretation of motion of three vortices on
a sphere, that was used in~\cite {BorLeb, BorMam}. In Fig. 1 we
represent the surface of integral
$$
\boldsymbol{D}=a_1M_1+a_2M_2+a_3M_3
$$

\noindent
in the space of~$M_1,\,M_2,\,M_3$, where~$a_i=\frac{1}{\Gamma_i}$~are inverse
intensities of vortices, and~$M_i$~are squared distances between
vortices. The triangular area presented in Fig.~1 is selected by
the conditions~$M_i\ge0$.

\noindent
We have painted the areas of nonphysical states, for which
the triangle inequality is not fulfilled over, in black. The curves represent
levels of the energy
integral. Thomson's configuration on this figure corresponds to the
point~$A$, where~$M_1=M_2=M_3$. As we can see from the
figure, under small perturbation of Thomson's configuration the system
starts  moving along the level of the energy integral and $D$, close to the point
$A$, which is bounded. Thus, after perturbation the system moves in a small
neighbourhood of the point $A$, without leaving it. Analogous arguments can be
carried out for the case of a plane.
Hence, Thomson's configurations of three vortices are ``relatively''
Lyapunov stable both on a sphere and on a plane.

\subsection*{The case of $\boldsymbol{ N=4,\,5,\,6;\,\boldsymbol\theta_0\in[0,\,\boldsymbol\theta_0^*)}$}

To show the ``relative'' Lyapunov stability  of Thomson's configurations
for~$N=4,\,5,\,6$ and for ${\theta_0\in[0,\,\theta_0^*)}$ let us
diagonalize the quadratic
part of the Hamiltonian~$H_2$. Such
a diagonalization, as it has been already said above, is carried out with the help of
Fourier transformation given by the matrix
$$
T_{kl}=\frac{1}{\sqrt{N}}e^{-\frac{2\pi i}{N}kl}\,.
$$
After this diagonalization and elimination of a degree of freedom,
corresponding to the motion of vortices as a solid-state configuration,
we obtain:
\begin{equation}
\label{f11}
\begin{aligned}
H_2^{(4)}  ={}&\frac{3\Gamma^2}{4\pi}\cot\theta_0(p_1^2+q_1^2+p_2^2+q_2^2)+
\frac{\Gamma^2}{2\pi}\frac{\sqrt{3\cos^2\theta_0-1}}{\sin\theta_0}(p_3^2+q_3^2)\,,\\
H_2^{(5)}  ={}&\frac{\Gamma^2}{\pi}\cot\theta_0(p_1^2+q_1^2+p_2^2+q_2^2)+
\frac{\Gamma^2}{2\pi}\frac{\sqrt{3(2\cos^2\theta_0-1)}}{\sin\theta_0}
(p_3^2+q_3^2+p_4^2+q_4^2) \,,\\
H_2^{(6)}  ={}& \frac{5\Gamma^2}{4\pi}\cot\theta_0(p_1^2+q_1^2+p_2^2+q_2^2)+
\frac{3\Gamma^2}{4\pi}\frac{\sqrt{5\cos^2\theta_0-4}}{\sin\theta_0}(p_5^2+q_5^2)+\\
{} & +\frac{\Gamma^2}{2\pi}\frac{\sqrt{2(5\cos^2\theta_0-3)}}{\sin\theta_0}
(p_3^2+q_3^2+p_4^2+q_4^2)  \,.
\end{aligned}
\end{equation}

For~$\theta_0\in(0,\,\theta_0^*)$ the quadratic Hamiltonians~(\ref{f11}) are
positively defined and, consequently, can be used as Lyapunov functions. Thus,
Thomson's configurations of four, five and six vortices are ``relatively''
 Lyapunov stable on a sphere in areas of their linear stability. Proceeding
to limit~$\theta_0\to0$, it is also possible to
show a ``relative'' Lyapunov  stability on a plane.

\section{The case $\boldsymbol{N=7}$ on a plane}

Thomson's configuration are unstable in linear
approximation on a sphere in case $N=7$. On a plane the linear instability vanishes,
but appears in the system
of two zero frequencies with two-dimensional Jordan cells (the neutral case of
a indifferent equilibrium). After diagonalization the quadratic
Hamiltonian of the system takes the form
\begin{equation}
\label{f12}
H_2=\frac{a}{2}(q_1^2+q_2^2)+\frac{\omega_1}{2}(p_3^2+q_3^2+p_4^2+q_4^2)+
\frac{\omega_2}{2}(p_5^2+q_5^2+p_6^2+q_6^2)\,,
\end{equation}
where
$a=\frac{3}{\pi},\,\omega_1=\frac{\sqrt{5}}{4\pi},\,\omega_2=\frac{3}{4\pi}$,
and the values of $\Gamma$ and $R$ are assumed to be 1. It can be seen that this
system has two resonances of type 1:1 (variables 3, 4 and 5, 6) and two
zero frequencies with two-dimensional Jordan cells (variables 1 and 2). The
Hamiltonian~(\ref{f12}) is not positively defined, and, therefore, to study
its stability it is necessary to investigate expansions of the Hamiltonian of higher
order. For simplification of the Hamiltonian form we shall carry out the
procedure of Deprit-Hori normalization (a detailed exposition of the procedure
of Deprit-Hori normalization with the presence of zero frequencies with
two-dimensional Jordan cells is presented in Appendix). After normalization
the Hamiltonian of the third degree will have the form:
\begin{equation}
\label{f13}
H_3=b(\frac{1}{2}(p_1-p_2)(p_4^2+q_4^2-p_3^2-q_3^2)+(p_1+p_2)(p_3p_4+q_3q_4))\,,
\end{equation}
where~$b=0.0672552$. $ H_3 $ does not depend on the
variables~$p_5,\,q_5,\,p_6,\,q_6$. Therefore, degrees of freedom,
corresponding to these variables in approximation of the third order, are
separated. Further we can consider a system with four degrees of
freedom only.

To prove the instability of this system let us use a method of
constructions of quasihomogeneous truncations of expansion of a
Hamiltonian~\cite{Furta}. Thus for a system obtained from the initial one by
quasihomogeneous truncation, we try to find sets of partial solutions as
twisted rays~$\sim\frac{1}{t^\beta}\sin(\omega\ln t+\alpha)$. We have
considered basic quasihomogeneous truncations of a system of the third
degree. However, non of them has a partial solution in form of twisted
rays.

The book~\cite {Furta} presents constructions of increasing solutions for
systems with two degrees of freedoms with resonances 1:1 and two zero
frequencies with two-dimensional Jordan cells. In the case of  resonance 1:1,
to construct the increasing solutions we must have the terms of no less than
the~$4^{th}$ degree in variables, corresponding to the resonance; and in case
of two zero frequencies with two-dimensional Jordan cells the terms must be
no less than of the third degree.  Due to particularity of the system
and a large number of degrees of freedom, there is no such terms in this case,
$H_3$ includes only ``mixed'' terms (depending both on variables,
corresponding to
the resonance 1:1, and on variables, corresponding to zero frequencies).
Probably, this is the reason why it is impossible to find
increasing partial solutions.

Motion equations of the system of the third degree have an invariant
manifold
\begin{equation}
\label{f14}
p_2=q_2=p_3=q_3=p_4=q_4=p_5=q_5=p_6=q_6=0\,.
\end{equation}
The equations of motion on this manifold take the form:
$$
\left\{
\begin{aligned}
\dot p_1 & = -aq_1\,, \\
\dot q_1 & = 0\,.
\end{aligned}
\right.
$$
A solution of the given system is a set of particular solutions, which
increase linearly with time. However, a linearly increasing solution of the
truncated system can not be extended to an asymptotically increasing solution
of the complete system (which is possible for solutions in the form of twisted rays).
Therefore, we shall consider the further expansion of the Hamiltonian up to
terms of the fourth degree.

The normal form of the 4-th degree can be written as
\begin{equation}
\label{f15}
\begin{aligned}
H_4  ={}& 0.943361q_1p_1q_2p_2+0.728898W_{12}R_{56}-0.518170W_{34}R_{56}+ \\
{} & +0.010176R_{56}^2-0.518170W_{34}W_{56}+0.378841R_{34}R_{56}+ \\
{} & +0.003197R_{12}^2+0.513722W_{12}W_{34}-0.120745R_{12}R_{34}+ \\
{} & +0.005339W_{34}^2-0.76260W_{34}R_{34}-0.559242W_{12}R_{34}- \\
{} & -0.080602W_{34}R_{12}+0.353421W_{56}R_{34}+0.237601W_{12}^2- \\
{} & -0.013751W_{56}R_{56}-0.006875W_{56}^2+0.769745W_{12}W_{56}- \\
{} & -0.030918R_{34}^2\,,
\end{aligned}
\end{equation}
where
$$
\begin{aligned}
W_{12}={}&p_1q_2-p_2q_1\,,{}& \qquad
R_{12}={}&p_1^2+p_2^2\,, \\
W_{34}={}&p_3q_4-p_4q_3\,, {}& \qquad
R_{34}={}&\frac{1}{2}(p_3^2+q_3^2+p_4^2+q_4^2)\,, \\
W_{56}={}&p_5q_6-p_6q_5\,,{}& \qquad
R_{56}={}&\frac{1}{2}(p_5^2+q_5^2+p_6^2+q_6^2)\,.
\end{aligned}
$$
In the case of~$H_4$, an analogy with~$H_3$,  we did not manage to find a
quasihomogeneous truncation of the system which would admit increasing partial
solutions in the form of twisted rays.

Taking into account terms of the fourth degree in the Hamiltonian, equations on
the manifold~(\ref{f14}) have the form:
\begin{equation}
\label{f16}
\left\{
\begin{aligned}
\dot p_1 & =-aq_1\,, \\
\dot q_1 & =kp_1^3\,,
\end{aligned}
\right.
\end{equation}
where~$k=0.127892$. The equations~(\ref{f16}) have an integral
$$
\frac{kp_1^4}{4}+\frac{aq_1^2}{2}=const\,.
$$
Levels of the integral on a phase plane~$p_1,\,q_1$ represent
closed oval curves of the fourth degree with their center in the origin.
Hence, motions of the system on the manifold~(\ref{f14}) will be
limited. Thus, the  addition of terms of the fourth degree in the Hamiltonian has
resulted in smoothing of linear instability, appearing in analysis of the
Hamiltonian of the third degree.

The further normalization is connected with large computing complexity and,
strictly speaking, the problem of stability for~$N=7$ remains open.
This problem, in a sense, is a limiting problem for systems, containing
a parameter, which is the curvature in the problem under consideration.
For the zero
value of the parameter a linear analysis is insufficient, and we have to
take into account the nonlinear terms. In favour of instability we can
specify the fact that, with adding of small curvature, the solutions become
unstable in linear approximation. However, this reason cannot be considered
as a strict proof of instability.

\section{Conclusion}

Thus, the following generalization of Thomson's theorem on stability of
regular $N$-gon configurations of vortices on a sphere is true:

{\bf Theorem 1.} {\it
Let us consider regular $N$-gon configurations of vortices on a sphere,
located on the same latitude~$\theta_0$\/$:$}

$1.$ {\it For~$N\ge7$, they are unstable with respect to
deformations\/$;$}

$2.$ {\it For~$N=6,\,5$ and~$4$, they are Lyapunov stable  with respect to
deformations for~$\theta_0\in(0,\,\theta_0^*(N))$, and unstable
for~$\theta_0\in(\theta_0^*(N),\,\frac{\pi}{2})$. The limiting
latitudes of the stability~$\theta_0^*(N)$ are determined by formulas}
$$
\theta_0^{*}(6)=\arccos\frac{2}{\sqrt{5}}\mbox{, }
\theta_0^{*}(5)=\frac{\pi}{4}\mbox{, }
\theta_0^{*}(4)=\arccos\frac{1}{\sqrt{3}};
$$

$3.$ {\it For~$N=2,\,3$, they are Lyapunov stable  with respect to deformations
for all~$\theta_0$.}

{\bf Remark 1.}  {\it
The problem of stability at~$\theta_0=\theta_0^*$ at~$N=4,\,5$ and 6
requires separate analysis with respect to nonlinear terms and remains unsolved
to the present day.}

\noindent As for a plane, we have the following:

{\bf Theorem 2.}  {\it
Regular $N$-gon configurations of vortices on a plane are Lyapunov stable
with respect to deformations for~$N<7$, and unstable for~$N\ge8$.
The problem of stability for~$N=7$ remains unsolved.}

\section{Appendix}
In this work we use the method of Deprit-Hori normalization for
construction of normal forms of Hamiltonian near a fixed point.
A detailed exposition of this method in case of absence of Jordan
cells can be found, for example, in~\cite{Markeev}.
Here we shall generalize this
algorithm to the case of the presence of two-dimensional Jordan cells. To
determine a normal form we shall apply the transformation close to
identical, defined by a Hamiltonian~$W$. The Hamilton form of
transformation ensures its canonical properties. The initial
Hamiltonian~$H$, a normalized Hamiltonian~$K$ and the function~$W$
are expanded into a series in terms of a small parameter near a fixed point
$$
\begin{aligned}
 H  = \sum_{n=0}^\infty H_n\,, \qquad
 K  = \sum_{n=0}^\infty K_n\,, \qquad
 W  = \sum_{n=1}^\infty W_n\,.
\end{aligned}
$$
Here all functions~$H_n,\,K_n$ and~$W_n$ are polynomials of~$n$ degree
with respect to phase variables.
A~normalized Hamiltonian up to~$k$ degree is obtained by successive
solution of operator equations
\begin{equation}
\label{s1}
\begin{aligned}
 K_0 & =H_0\,, \\
 K_n & =H_n+\sum_{m=0}^{n-2}\Bigl[L_{m+1}H_{n-m-1}+\frac{1}{(n-m)!}
H_n^{(n-m)}\Bigr]+L_nH_0\mbox{,  } n\ge1\,,
\end{aligned}
\end{equation}
where
$$
H_m^{(i)}=\sum_{j=0}^{m}L_{j+1}H_{m-j}^{(i-1)}\mbox{, } H_m^{(0)}=H_m
$$
for~$n=0,\,1,\,\ldots,\,k$.

Thus, on the~$n$-th step, the normalization procedure is reduced to the solution
of the equation
\begin{equation}
\label{s2}
K_n=H^{kn}_n+L_nH_0\,,
\end{equation}
where~$H^{kn}_n$~is a known function, depending on functions already found
on the previous steps of normalization.

As we consider the normalization near a fixed point, it is obvious that
the expansion
of~$H$ will begin with the second order terms. Hence, the corresponding
expansions of~$K$ and~$W$ will begin with the second and the third order terms.
The equation~(\ref{s2}) can be rewritten in the form
\begin{equation}
\label{s3}
K_n=H_n^{kn}+\mathcal{D} W_n,
\end{equation}
where~$\mathcal{D} f=\{H_2,\,f\}$.

While solving the equation~(\ref{s3}), the function~$W_n$ is
selected so that all terms in the right part of the equation~(\ref{s3}) are
excluded, except those, which could not be presented as~$M=\mathcal{D} f$,
where~$f$~is some function. $K_n$ will consist of these remained
terms.

\subsection{Absence of Jordan cells}

Let a quadratic part of the Hamiltonian~$H_2$ lack Jordan cells.
Then after diagonalization and complex change
$$
 z=p+iq\,, \qquad
 \bar{z}=p-iq\,, \qquad
 \{z_k, \bar{z_l}\}=\frac{1}{2i}\delta_{kl}
$$
the quadratic part of the Hamiltonian~$H_2$ has the form:
$$
 H_2=\sum_i\omega_iz_i\bar{z_i}.
$$
The action of an operator~$\mathcal{D}$ on some
monomial~$M_{\boldsymbol l,\boldsymbol m}=A_{\boldsymbol l,\boldsymbol m}z^l\bar{z}^m$ is defined by expression
\begin{equation}
\label{s4} \mathcal{D} M_{l,m}=\frac{1}{2i}{\boldsymbol\omega}
(\boldsymbol{m}-\boldsymbol{l})M_{\boldsymbol l,\boldsymbol
m}=C_{\boldsymbol l,\boldsymbol m}M_{\boldsymbol l,\boldsymbol m},
\end{equation}
where~$\boldsymbol\omega$~is a vector of frequencies. The operator~$\mathcal{D}$ does not
change the degree of monomial, but only multiplies it by some constant.

Thus, in the right part of~(\ref{s1}) we are unable to eliminate only those
monomials, for which the following equality
\begin{equation}
\label{s5}
C_{\boldsymbol l,\boldsymbol m}=-\frac{1}{2i}{\boldsymbol\omega}(\boldsymbol m-\boldsymbol l)=0
\end{equation}
is fulfilled.
A normalizing function and a normalized Hamiltonian will take the form:
$$
\begin{aligned}
 K_n  = \sum_{{l,m}\atop{C_{l,m}=0}}M_{\boldsymbol l,\boldsymbol m}\,, \qquad
 W_n  = \sum_{{l,m}\atop{C_{l,m}\ne0}}\frac{M_{\boldsymbol l,\boldsymbol m}}{C_{\boldsymbol l,\boldsymbol m}}\,,
 \end{aligned}
$$
where~$M_{\boldsymbol l,\boldsymbol m}$~are monomials included in~$H_n^{kn}$.

\subsection{A case of two-dimensional Jordan cells}

Let~$H_2$ have~$k$ zero frequencies with two-dimensional Jordan
cells, then~$H_2$ can be presented as
$$
H_2=\sum_{i=1}^{k}q^2_i+\sum_{i=k+1}^N\omega_i(q^2_i+p^2_i)\,.
$$
After a complexification with respect to
variables~$q_{k+1},\,\ldots,\,q_N^{\vphantom{9}},
\,p_{k+1},\,\ldots,\,p_N^{\vphantom{9}}$ we  obtain
$$
H_2=\sum_{i=1}^{k}q^2_i+\sum_{i=1}^{N-k}\omega_iz_i\bar{z_i}
=\sum_{i=1}^{k}q^2_i+H_2'\,.
$$
 In this case, the operator~$\mathfrak{D}$ has the form
\begin{equation}
\label{s6}
\mathfrak{D}=\mathfrak{D}'+\sum_{i=1}^{k}2q_i\frac{\partial}{\partial p_i}\,,
\end{equation}
where the operator~$\mathfrak{D}'$ acts on a space of
variables~$z_{1},\,\ldots,\,z_{N-k}^{\vphantom{9}},\,\bar {z_{1}},\,\ldots,
\,\bar z_{N-k}^{\vphantom{9}}$,
and has the same properties, as the operator~$\mathfrak{D}$ in Section 5.1 (i.\,e. it does not
change the degrees of monomials). In this case, the operator~$\mathfrak{D}$ does not
already preserve
the form of monomials. Let us consider its action on some monomial
$$
M_{\boldsymbol l,\boldsymbol m}=M_{\boldsymbol s,\boldsymbol r}'p_{1}^{l_1}\ldots p_k^{l_k}
q_1^{m_1}\ldots q_k^{m_k},
$$
where
$$
M_{\boldsymbol s,\boldsymbol r}'=z_1^{s_1}\ldots z_{N-k}^{s_{N-k}^{\vphantom{9}}}
\bar{z_1}^{r_1}\ldots\bar{z}_{N-k}^{m_{N-k}^{\vphantom{9}}}.
$$

\noindent
Applying the operator~$\mathfrak{D}$ to~$M_{\boldsymbol l,\boldsymbol m}$ we obtain
\begin{equation}
\label{s7}
\mathfrak{D} M_{\boldsymbol l,\boldsymbol m}=C_{\boldsymbol s,\boldsymbol
r}M_{\boldsymbol l,\boldsymbol m}+M_{\boldsymbol l,\boldsymbol
m}\sum_{i=1}^{k} 2l_i\frac{q_i}{p_i}\,, \qquad
C_{\boldsymbol s,\boldsymbol r}=\mathfrak{D}'M_{s,r}'=\frac{1}{2i}\boldsymbol\omega(\boldsymbol
r-\boldsymbol s)\,.
\end{equation}
Let
$$
H_n^{kn}=\sum_{\boldsymbol l,\boldsymbol m}A_{\boldsymbol l,\boldsymbol m}^{(n)}M_{\boldsymbol l,\boldsymbol
m}\mbox{, }
W_n=\sum_{\boldsymbol l,\boldsymbol m}B_{\boldsymbol l,\boldsymbol m}^{(n)}M_{\boldsymbol l,\boldsymbol
m}\mbox{, }
K_n=\sum_{\boldsymbol l,\boldsymbol m}K_{\boldsymbol l,\boldsymbol m}^{(n)}M_{\boldsymbol l,\boldsymbol m}\,,
$$
then the equation~(\ref{s3}) for coefficients of monomial~$M_{\boldsymbol l,\boldsymbol m}$ will
take the form
\begin{equation}
\label{s8} K_{\boldsymbol l,\boldsymbol m}^{(n)}=A^{(n)}_{\boldsymbol
l,\boldsymbol m}+B^{(n)}_{\boldsymbol l,\boldsymbol m}C_{\boldsymbol
s,\boldsymbol r}+ \sum_{i=1}^kB^{(n)}_{\boldsymbol l+ \boldsymbol {e_i, m-e_i}}2(l_i+1)\,.
\end{equation}
Here~$\boldsymbol {e_i}$~is unit vectors of an integer lattice. From~(\ref{s8}) it is clear
 that it is impossible to set~$K^{(n)}_{\boldsymbol l,\boldsymbol m}$ equal to zero,
if both conditions are fulfilled simultaneously:

1) $C_{\boldsymbol s,\boldsymbol r}=0$ i.\,e. $M'_{\boldsymbol s,\boldsymbol r}$~is a resonance monomial;

2) $m_i=0,\,i=1,\,\ldots, k$, thus the sum in~(\ref{s8}) is absent, since for
negative components~$\boldsymbol l, \boldsymbol m$~$M^{(n)}_{\boldsymbol l,\boldsymbol m}$ do not exist.

The monomials which satisfy these two conditions are resonance monomials.
However, as we shall see further, there are also other resonance monomials.

Let us obtain now a normalizing function, which will help us to exclude nonresonance
monomials. Let us consider separately 2 cases:

I. $M_{\boldsymbol s,\boldsymbol r}'$ is not resonance;

II. $ M_{\boldsymbol s,\boldsymbol r}'$ is resonance, but $m_i\ne0$.

{\bf Case I.} Let us consider an integer lattice~$l_1,\,\ldots,\,l_k$. For
each monomial~$M_{\boldsymbol l,\boldsymbol m}$ and
function~$w_{\boldsymbol l,\boldsymbol m}$, which excludes the latter, we
shall put in the correspondence a point of this lattice. Let us
denote~$L$-plane as the hyperplane cutting off a $k$-dimensional angle in
$\boldsymbol l$-space, determined by the equation $$
\sum\limits_{i=1}^kl_i=L\,.
$$
Let~$H_n^{kn}$ contain a monomial~$M_{\boldsymbol l,\boldsymbol m}$. The
given monomial can be excluded from the right part of~(\ref{s3}) with the
help of function~$w_{\boldsymbol l,\boldsymbol m}=-\frac{M_{\boldsymbol
l,\boldsymbol m}}{C_{\boldsymbol s,\boldsymbol r}}$. Under this
elimination, as we can see from~(\ref{s7}), in the right part
of~(\ref{s3}) we get the sum of monomials
\begin{equation}
\label{s9}
\sum_{i=1}^k\biggl(-\frac{2l_i}{C_{\boldsymbol s,\boldsymbol r}}\biggr)
M_{\boldsymbol l-\boldsymbol {e_i},\boldsymbol{m+e_i}}\,.
\end{equation}
Thus,  we pass from the point~$\boldsymbol l$ to the
neighbouring points~$\boldsymbol l-\boldsymbol{e_i},\,i=1,\,\ldots,\,k$. Such passage is always
made from~$L$-plane into~$(L-1)$-plane, and the monomials are multiplied
by~$\biggl(-\frac{2l_i}{C_{\boldsymbol s,\boldsymbol r}}\biggr)$. Using the same method we
exclude originating monomials and move along a lattice in the direction to
the origin. In the general case an arbitrary point of the lattice can be
reached by several paths, and at the same time appearing monomials are
summarized. When any coordinate  becomes equal to zero at such a motion,
the further motion takes place in remaining variables only. In  Fig.\,2
the lattice and such passages at $k=2$ are presented. A sequence of the
passages constructed in such a way breaks on a point~$\boldsymbol l=\boldsymbol 0$.

Adding normalizing functions~$w_{\boldsymbol l,\boldsymbol m}$ by all
points of a rectangle, we obtain a function~$W_{\boldsymbol l,\boldsymbol m}$, that excludes a
monomial~$M_{\boldsymbol l,\boldsymbol m}$:
\begin{equation}
\label{s10}
W_{\boldsymbol l,\boldsymbol m}=-\frac{1}{C_{\boldsymbol s,\boldsymbol r}}\sum_{\boldsymbol l'}M_{\boldsymbol l',\boldsymbol m'}
\frac{l_1!\,l_2!\ldots l_k!}{l'_1!\,l'_2!\ldots l'_k!}
\biggl(-\frac{2}{C_{\boldsymbol s,\boldsymbol r}}\biggr)^{l_1-l'_1+l_2-l'_2+\ldots +l_k-l'_k}
f_{\boldsymbol l',\boldsymbol l},
\end{equation}
where~$\boldsymbol {m'}=\boldsymbol m+(\boldsymbol l-\boldsymbol {l'})$,
$f_{\boldsymbol l',\boldsymbol l}$~is a number of paths from point~$\boldsymbol l$
into point~$\boldsymbol {l'}$ moving in direction of origin. This number can  be
calculated with the help of recurrence relation
$$
f_{\boldsymbol {l'},\boldsymbol l}=\sum_{i=1}^kf_{\boldsymbol {l'}+\boldsymbol {e_i},\boldsymbol l}(1-\delta_{l'_i,l_i})\,.
$$
Besides the successive elimination of monomials of the first type with the
help of functions~(\ref{s10}), they can also be eliminated with the help
of functions~$w_{\boldsymbol l,\boldsymbol m}$. For this method it is neccessary to take such
order of enumeration of $\omega_{l,m}$ that with elimination of the next
monomial in the right part of~(\ref{s3}), the excluded earlier monomials
do not appeared again. Since the elimination of monomial~$M_{\boldsymbol l,\boldsymbol m}$
with the help of~$w_{\boldsymbol l,\boldsymbol m}$ we pass from~$L$-plane
into~$(L-1)$-plane, such order is a sequential enumeration of all
monomials with maximum~$L=n$, then with~$L$ smaller on one unit, etc.

{\bf Case II.}
In this case an action of the operator~$\mathcal D$ over monomial~$M_{\boldsymbol l,\boldsymbol m}$ has the
form
\begin{equation}
\label{s11}
\mathcal D M_{\boldsymbol l,\boldsymbol m}=\sum_{i=1}^k2l_iM_{\boldsymbol l-\boldsymbol {e_i},\boldsymbol m+\boldsymbol{e_i}}\,.
\end{equation}
Thus, it is possible to exclude the monomial~$M_{\boldsymbol l,\boldsymbol m}$ with the help of the
functions
\begin{equation}
\label{s12}
w_{\boldsymbol l,\boldsymbol m}^{(i)}=-\frac{1}{2(l_i+1)}M_{\boldsymbol l+\boldsymbol {e_i},\boldsymbol m-\boldsymbol {e_i}}.
\end{equation}
And in the right part of~(\ref{s3}) the sum
\begin{equation}
\label{s13}
\sum_{j\ne i}\biggl(-\frac{l_j}{l_i+1}\biggr)
M_{\boldsymbol{l+e_i-e_j,m-e_i+e_j}}
\end{equation}
is obtained.
As we can see from the interpretation of an integer lattice, with such
elimination we pass from the point~$\boldsymbol l$ in
points~$\boldsymbol{l+e_i-e_j},\,j=1,\,\ldots, k,\,j\ne i$. Thus there occures a motion in
one~$L$-plane.

Successively applying transformation with the functions~(\ref{s12}) for
various~$ i $ we obtain some chain of monomials. Selecting the coefficients
for excluding functions~(\ref{s12}) we can reduce the whole chain to any
of its monomials. Thus, if we can exclude only one of monomials from a chain
without emerging of new terms, the whole chain can be excluded from the
right part of~(\ref{s3}). The monomial can be completely excluded, when
all terms in the sum~(\ref{s13})  equal zero, i.\,e. when for some
of~$i$ we have~$l_j=0,\,j\ne i$ and~$m_i>0$. We can reduce
an arbitrary monomial to such form (and exclude it completely), if there
exists at least one $i$,
 such that
\begin{equation}
\label{s14}
m_i>\sum_{j\ne i}l_j\,.
\end{equation}
We can prove it with the use of an invariance of the inequality~(\ref{s14})
with respect to  the
transformations~(\ref{s12}). If for the monomial~$M_{\boldsymbol {l, m}}$ the
inequalities
\begin{equation}
\label{s15}
m_i\le\sum_{j\ne i}l_j\,, \quad i=1,\,\dots, k
\end{equation}
are fulfilled,
the given monomial cannot be excluded completely from the right part. We
can only transfer it into any of monomials from the chain. Thus monomials,
satisfying~(\ref{s15}) are resonance monomials.

Let us consider the case, when for some~$i$ the inequality (\ref{s14}) is
fulfilled, i.\,e. the monomial can be excluded completely. Let us construct a
function~$W_{\boldsymbol {l, m}}$ while $h$ will exclude the monomial.
After an elimination of this monomial with the
help of the functions~(\ref{s12}) we pass from point~$\boldsymbol l$ into
points~$\boldsymbol {l+e_i-e_j},\,j\ne i$. On $L$-plane the motion takes place in
direction to an axis~$l_i$. At~$k=3$ such a plane and the sequence
of passages are represented in Fig.~3. For each passage of monomial from
some point into another one, the monomial is multiplied by~$-\frac{l_j}{l_i+1}$.
As in case I, summing up~$w_{\boldsymbol {l, m}}$ by~$(k-1)$-dimensional
rectangle, we obtain
\begin{equation}
\label{s16}
W_{\boldsymbol {l, m}}=\sum_{l'_1=0}^{l_1}\ldots\sum_{l'_{i-1}=0}^{l_{i-1}}
\sum_{l'_{i+1}=0}^{l_{i+1}}\ldots\sum_{l'_k=0}^{l_k}
(-1)^{l'_i-l_i+1}\frac{l_1!\,l_2!\ldots l_k!}{l'_1!\,l'_2!\ldots l'_k!}
\frac{f_{\boldsymbol {l', l}}}{2(l'_i+1)}M_{\boldsymbol{l'+e_i, m'-e_i}}\,,
\end{equation}
where
$$
l'_i  =l_i+\sum_{j\ne i}(l_j-l'_j),\quad
m'_j  =m_j+(l_j-l'_j), \quad i=1,\,\ldots, k\quad
f_{\boldsymbol {l', l}}  = \sum_{j\ne i}f_{\boldsymbol {l'-e_i+e_j,l}}
(1-\delta_{l'_j,l_j})\,.
$$

 Thus, in the case of~$k$ zero frequencies with two-dimensional Jordan
cells normal form of the Hamiltonian include the
monomials~$M_{\boldsymbol {l, m}}=M_{\boldsymbol {s, r}}'p_1^{l_1}\ldots p_k^{l_k}q_1^{m_1}\ldots
q_k^{m_k}$ such that~$M_{\boldsymbol {s, r}}'$ are resonance and the following inequality
$$
m_i\le\sum_{j\ne i}l_j\,, \quad\mbox{ for }\forall\,i=1,\,\ldots,k
$$
is fulfilled.

\begin{figure}[ht!]
$$
\includegraphics{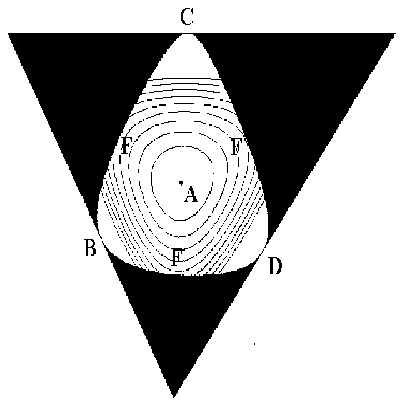}
$$
\caption{}
\end{figure}

\begin{figure}[ht!]
$$
\includegraphics{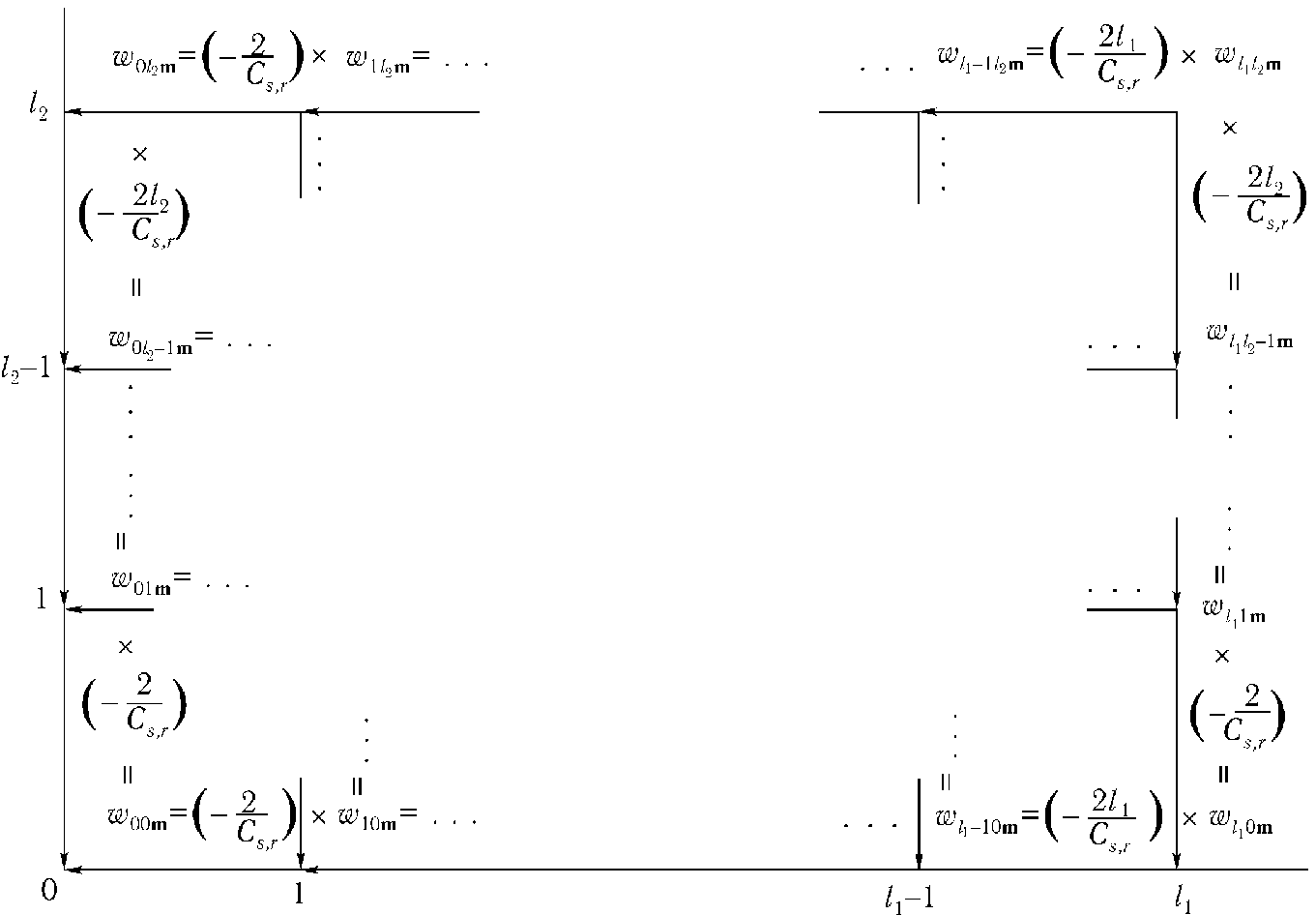}
$$
\caption{}
\end{figure}

\begin{figure}[ht!]
$$
\includegraphics{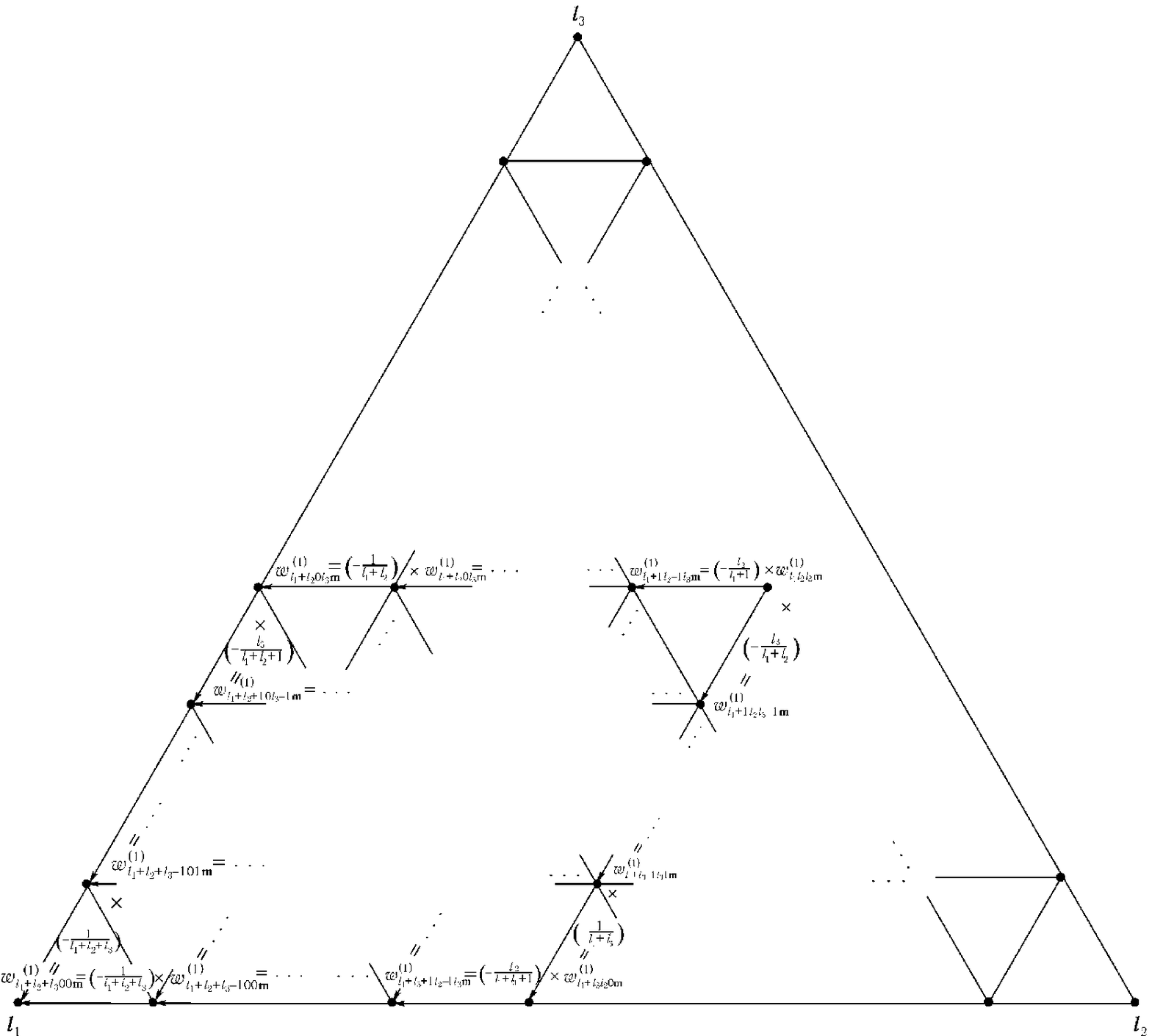}
$$
\caption{}
\end{figure}

\end{document}